\documentclass[conference]{IEEEtran}
\IEEEoverridecommandlockouts
\usepackage[detect-all,group-separator={,}]{siunitx}
\usepackage{amsmath,amssymb,amsfonts}
\usepackage{balance}
\usepackage{booktabs}
\usepackage{cite}
\usepackage{comment}
\usepackage{csquotes}
\usepackage{algorithmic}
\usepackage[hidelinks]{hyperref}
\usepackage{inconsolata}
\usepackage[inline]{enumitem}
\usepackage{graphicx}
\usepackage{lipsum}
\usepackage{multicol}
\usepackage{multirow}
\usepackage{textcomp}
\usepackage{lipsum}
\usepackage[dvipsnames]{xcolor}

\newif\ifdraft
\draftfalse

\def\BibTeX{{\rm B\kern-.05em{\sc i\kern-.025em b}\kern-.08em
    T\kern-.1667em\lower.7ex\hbox{E}\kern-.125emX}}

\newcommand{\mofa}{\texttt{MOFA}}
\newcommand{\difflinker}{\texttt{DiffLinker}}
\newcommand{\moflinker}{\texttt{MOFLinker}}
\newcommand{\parsl}{\texttt{Parsl}}

\newcommand{\Inputs}{\mathcal{I}}

\ifdraft
  \newcommand{\logan}[1]{{\textcolor{teal}{\textbf{Logan:}~\enquote{#1}}}}
  \newcommand{\eliu}[1]{{\textcolor{cyan}{\textbf{Eliu:}~\enquote{#1}}}}
  \newcommand{\ian}[1]{{\textcolor{orange}{\textbf{Ian:}~\enquote{#1}}}}
  \newcommand{\greg}[1]{{\textcolor{purple}{\textbf{Greg:}~\enquote{#1}}}}
  \newcommand{\nathaniel}[1]{{\textcolor{blue}{\textbf{Nathaniel:}~\enquote{#1}}}}
  \newcommand{\daniel}[1]{{\textcolor{Emerald}{\textbf{Daniel:}~\enquote{#1}}}}
  \newcommand{\marcus}[1]{{\textcolor{ForestGreen}{\textbf{Marcus:}~\enquote{#1}}}}
  \newcommand{\haochen}[1]{{\textcolor{CadetBlue}{\textbf{Haochen:}~\enquote{#1}}}}
  \newcommand{\xiaoli}[1]{{\textcolor{Mulberry}{\textbf{Xiaoli:}~\enquote{#1}}}}
  \newcommand{\hassan}[1]{{\textcolor{Mahogany}{\textbf{Hassan:}~\enquote{#1}}}}

  \newcommand{\fm}[1]{{\textcolor{red}
  {\textbf{FIXME:}~\enquote{#1}}}}
\else
  \newcommand{\todo}[1]{}
  \newcommand{\logan}[1]{}
  \newcommand{\eliu}[1]{}
  \newcommand{\daniel}[1]{}
  \newcommand{\ian}[1]{}
  \newcommand{\marcus}[1]{}
  \newcommand{\haochen}[1]{}
  \newcommand{\xiaoli}[1]{}
  \newcommand{\hassan}[1]{}
  \newcommand{\nathaniel}[1]{}
  \newcommand{\greg}[1]{}
  \newcommand{\fm}[1]{}
\fi

\def\equationautorefname~#1\null{Eq.~(#1)\null}
    
\begin{document}

\title{%
    \texttt{MOFA}: 
    Discovering Materials for Carbon Capture 
    with a GenAI- and Simulation-Based Workflow
}

\ifdraft
    \author{Anonymous Author(s)}
\else
    \author{
    \IEEEauthorblockN{
    	Xiaoli Yan\IEEEauthorrefmark{1}\IEEEauthorrefmark{2}\textsuperscript{\#}, 
        Nathaniel Hudson\IEEEauthorrefmark{1}\IEEEauthorrefmark{3}\textsuperscript{\#}, 
        Hyun Park\IEEEauthorrefmark{1}\IEEEauthorrefmark{4}, 
        Daniel Grzenda\IEEEauthorrefmark{3}, 
        J. Gregory Pauloski\IEEEauthorrefmark{3}, 
        Marcus Schwarting\IEEEauthorrefmark{3}, 
        \\
        Haochen Pan\IEEEauthorrefmark{3},
        Hassan Harb\IEEEauthorrefmark{1}, 
        Samuel Foreman\IEEEauthorrefmark{1}, 
        Chris Knight\IEEEauthorrefmark{1}, 
        Tom Gibbs\IEEEauthorrefmark{5}, 
        Kyle Chard\IEEEauthorrefmark{1}\IEEEauthorrefmark{3}, 
        Santanu Chaudhuri\IEEEauthorrefmark{1}\IEEEauthorrefmark{2}, 
        \\
        Emad Tajkhorshid\IEEEauthorrefmark{4}, 
        Ian Foster\IEEEauthorrefmark{1}\IEEEauthorrefmark{3}, 
        Mohamad Moosavi\IEEEauthorrefmark{6}, 
        Logan Ward\IEEEauthorrefmark{1}, 
        E.~A. Huerta\IEEEauthorrefmark{1}\IEEEauthorrefmark{3}\IEEEauthorrefmark{4}
    }
    \IEEEauthorblockA{
        \IEEEauthorrefmark{1}%
        Argonne National Laboratory; Lemont, IL, United States
    }
    \IEEEauthorblockA{
        \IEEEauthorrefmark{2}%
        University of Illinois Chicago; Chicago, IL, United States
    }
    \IEEEauthorblockA{
        \IEEEauthorrefmark{3}%
        University of Chicago; Chicago, IL, United States
    }
    \IEEEauthorblockA{
        \IEEEauthorrefmark{4}%
        University of Illinois Urbana-Champaign; Urbana, IL, United States
    }
    \IEEEauthorblockA{
        \IEEEauthorrefmark{5}%
        NVIDIA Inc.; Santa Clara, CA, United States
    }
    \IEEEauthorblockA{
        \IEEEauthorrefmark{6}%
        University of Toronto; Toronto, Ontario
    }
}
\fi

\maketitle
\ifdraft
\else
    \begingroup\renewcommand\thefootnote{\#}
    \footnotetext{Equal contribution.}
    \endgroup
\fi

\begin{abstract}
We present \mofa{}, an open-source generative AI (GenAI) plus simulation workflow for high-throughput generation of metal-organic frameworks (MOFs) on large-scale high-performance computing (HPC) systems.
\mofa{} addresses key challenges in integrating GPU-accelerated computing for GPU-intensive GenAI tasks, including distributed training and inference, alongside CPU- and GPU-optimized tasks for screening and filtering AI-generated MOFs using molecular dynamics, density functional theory, and Monte Carlo simulations. 
These heterogeneous tasks are unified within an online learning framework that optimizes the utilization of available CPU and GPU resources across HPC systems. 
Performance metrics from a 450-node (14,400 AMD Zen 3 CPUs +  1800 NVIDIA A100 GPUs) supercomputer run demonstrate that \mofa{} achieves high-throughput generation of novel MOF structures, with CO$_2$ adsorption capacities ranking among the top 10 in the hypothetical MOF (hMOF) dataset. 
Furthermore, the production of 
high-quality MOFs exhibits a linear relationship with the number 
of nodes utilized. The modular architecture of \mofa{} will
facilitate its integration into other scientific applications that dynamically combine GenAI with large-scale 
simulations.
\end{abstract}



\section{Introduction}
\label{sec:intro}
Carbon dioxide (CO$_2$) has been identified as the primary contributor to the elevation of earth's atmospheric temperature\cite{hansen_climate_1981}. The combustion of fossil fuel is the major source of CO$_2$ emission. CO$_2$ emitted and absorbed by non-human activity can be a self-contained cycle. Developing new technologies to capture CO$_2$ from the human activity can be one of the most efficient solutions to mitigate the global climate change. 

\textit{Metal-Organic Frameworks}~(MOFs) are materials 
that have drawn much attention 
in the scientific community 
for their potential in numerous applications, including carbon capture~\cite{zhang2018ultrahigh, fernandez2014rapid}.
MOFs typically comprise two main types of components: 
\textit{(i)} 
an organic molecule (\enquote{linker} or \enquote{ligand}; synonymous) 
    and 
\textit{(ii)} an inorganic metal (\enquote{cluster}),
organized in a topology that can allow them, for example, to store gases like hydrogen or carbon dioxide~\cite{li2018recent}.
This ability makes MOFs exciting for applications other than carbon capture such as catalysis~\cite{hao2021recent}, drug discovery~\cite{lawson2021metal}, and luminescence sensing~\cite{mof_lumen}. 

\textit{Generative AI}~(GenAI) methods are by now widely used
to rapidly generate fluid text and high-resolution images. 
Image generative tools using denoising diffusion models (e.g., Stable Diffusion~\cite{rombach2022high}) can generate photo-realistic images in response to a text prompt.
In materials science, 
diffusion models can be trained instead to generate novel molecular structures with target chemical properties as prompts~\cite{jablonka2020big, kang2024chatmof, park2024generative}.
But there remain challenges regarding how to
\textit{(i)} intelligently traverse chemical space and 
\textit{(ii)} efficiently execute large-scale, high-throughput MOF generative workflows at scale.

A simple brute-force approach to creating many new MOF structures would be to combine metal nodes with organic linkers in different geometries exhaustively.
However, the chemical search space for MOFs is intractably large due to the many possible metal nodes, organic linkers, and pore geometries~\cite{moosavi2020understanding}.
The use of a generic chemical GenAI model to produce linkers is also infeasible, as the resulting molecules are not more likely to produce better MOFs than those found with brute-force searches.
We instead need a GenAI that is fine-tuned to generate those rare molecules that yield MOF structures with \textit{interesting} chemical properties---thus requiring fewer guesses to reach the same answer.

We approach this task of producing a GenAI model able to efficiently explore the large space of possible MOFs as follows.
First, we construct an initial GenAI model for MOF linker generation, \moflinker{}, by fine-tuning an existing model developed for drug discovery on subspaces of high-performing MOFs.
Then, we refine \moflinker{} over time by repeating steps: 
\textit{(i)}~\textbf{generate} new linkers with \moflinker{}, 
\textit{(ii)}~\textbf{assemble} new MOFs from generated linkers, 
\textit{(iii)}~\textbf{screen} assembled MOFs to eliminate non-promising linkers, and 
\textit{(iv)}~\textbf{retrain} \moflinker{} using performant linkers identified through screening with the goal of improving the quality of linkers produced.
The screening step, in particular, is crucial to the success of this workflow.
Screening combines simple structural tests of MOF feasibility with the use of expensive 
quantum chemistry tools (\texttt{CP2K}~\cite{Khne2020}, \texttt{LAMMPS}~\cite{lammps}, \texttt{RASPA}~\cite{raspa2016}) to compute important MOF properties such as stability and porosity. 

\begin{figure*}[!htbp]
    \centering
    \includegraphics[width=0.825\linewidth]{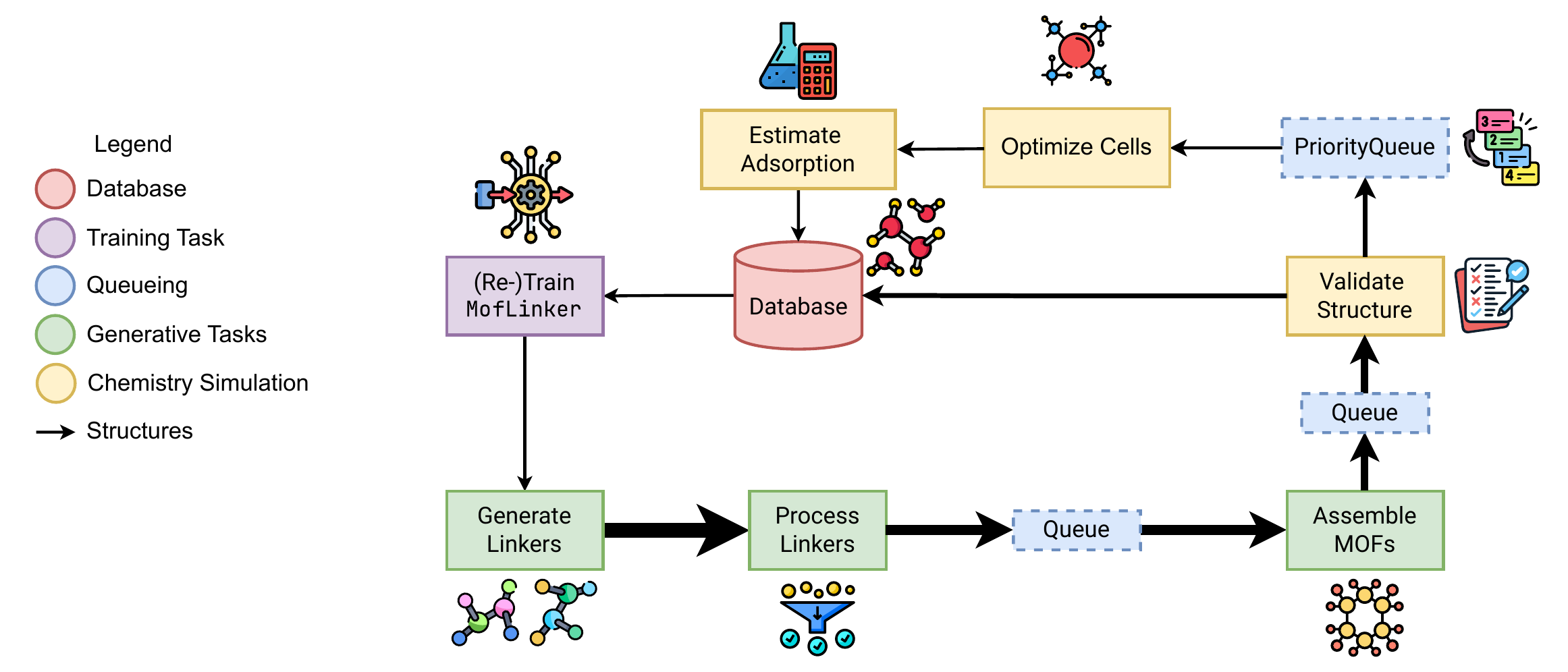}
    \caption{
        \mofa{} implements an online learning loop that 
        refines a generative AI model, \moflinker{}, 
        using the MOFs it has generated. 
        The initial steps in the workflow validate linker molecules produced by the generative model before using those that pass validation to assemble MOFs. 
        New MOFs are placed in a LIFO queue, from which they are retrieved to be evaluated for stability, 
        and the gas capacity of the most stable are further evaluated to refine the structures and estimate properties of interest. The structures and their computed properties are collected in a database and used to retrain the GenAI model. 
        All steps run concurrently.
        Note: The width of the arrows for \enquote{Structures} corresponds with the amount of structures being passed between each pair of tasks in the workflow.
    }
    \label{fig:workflow}
\end{figure*}

Our workflow thus combines GenAI, in the form of \moflinker{}, with both molecule assembly and a variety of screening computations to navigate chemical space. 
This combination of elements makes efficient workflow execution challenging.
First, massively parallel high-performance computing (HPC) is needed to enable rapid screening of many candidates molecules.
Second, tasks in the workflow have heterogeneous hardware requirements, with different tasks executing most efficiently on CPUs (e.g., molecule structure assembly), GPUs (e.g., GenAI training and inference), or a mix of both (e.g., quantum chemistry simulations).
This diversity poses a heterogeneous computing challenge: a high-throughput MOF generation workflow needs to create new and relevant MOF structures while coordinating the heterogeneous needs of the different task types.
Ideally, such a workflow will produce more novel, valid, and relevant MOF structures when consuming more node-hours on an HPC system.
%

To address these challenges, we introduce \mofa{},
an open-source computational workflow that uses online learning to generate novel MOF structures for carbon capture.%
\footnote{
    Repository for \mofa{} can be found here: 
    \texttt{https://github.com/XXX/YYY} (redacted to adhere to double-blind review).
}  
We focus the design of \mofa{} around the objective of generating MOFs for carbon capture.
At a high level, \mofa{} is a heterogeneous workflow consisting of two task types:
\begin{enumerate*}[label=\textit{(\roman*)}]
    \item GPU-optimized tasks for inference and training of a GenAI diffusion model for generating the building blocks of new MOFs
    and
    \item CPU- and GPU-optimized tasks that screen generated MOFs based on chemical properties calculated through atomistic simulations.
\end{enumerate*}
To dynamically schedule tasks with varying resource requirements, \mofa{} is built on the \texttt{Parsl}~\cite{babuji19parsl} and \texttt{Colmena} frameworks~\cite{ward2021colmena} to scale the workload across heterogeneous resources on large-scale computational systems.

The central contributions of this paper are as follows:
\begin{enumerate}
    
    \item 
    We describe an HPC-coupled-generative AI workflow, \mofa{}, 
    for high-throughput discovery of MOFs for carbon capture on heterogeneous HPC systems.
    Its open source, modular implementation will facilitate its use for both computer science research and other applications that combine GenAI and large scale simulations.
    
    \item 
    We demonstrate, via a 450-node (\num{14400} AMD Zen 3 CPUs + \num{1800} NVIDIA A100 GPUs), 3-hour run, that 
    \mofa{} can produce 114 MOFs per hour with competitive CO\textsubscript{2} adsorption capacities, with one MOF in the top 5 (4.05 mol/kg at 0.1 bar) and ten MOFs in the top 10\% (1--2 mol/kg at 0.1 bar) of the 4547-MOF structurally similar subset of the 137,652-MOF hMOF dataset~\cite{hmof_2012}.

    \item 
    We demonstrate in this and other runs, plus supporting analyses, that \mofa{} achieves high computational efficiency on modern GPU- and CPU-based HPC systems.

\end{enumerate}

\section{Related Work}
\label{sec:related}
Current approaches to accelerating the rational discovery of MOFs combine one or several of the following methods: AI, high throughput screening, and atomistic simulations.
Here, we discuss related work in MOF discovery and systems for executing heterogeneous computing workflows.

\subsection{MOFs \& Their Discovery}
\label{sec:mofs}

MOFs are versatile materials composed of metal ion clusters coordinated with organic linkers to form porous crystalline structures~\cite{wang2019state,mosca_stability_2023}.
Their tunable pore sizes, high surface areas, and structural flexibility have attracted significant attention since the 1990s, enabling them to serve in a range of applications such as gas storage, separation, and catalysis~\cite{fujita1994preparation, corma2010engineering, yang2017metal, chen2018metal}.

Prior to the adoption of GenAI approaches, MOF screening workflows often used brute-force~\cite{wilmer2012large,daglar2018computational}, heuristics-driven~\cite{fernandez2014rapid,fernandez2016geometrical}, or sampling strategies~\cite{mukherjee2022sequential}
to define and prioritize candidate MOF structures.
GenAI models, which are capable of producing novel experiments based on previous successful attempts, can augment these existing workflow strategies.
Model architectures such as diffusion, generative adversarial networks, and variational autoencoders have been employed in applications such as de novo drug screening \cite{swanson2024generative}, shape optimization \cite{li2020efficient}, chemical synthesis identification \cite{kim2017virtual}, and drug discovery~\cite{igashov2022equivariant}.

In the context of MOF design, \texttt{MOFDiff}~\cite{fu2023mofdiff} is a coarse-grained diffusion architecture developed to produce MOF structures with effective CO\textsubscript{2} separation capabilities.
\texttt{MOFDiff} starts with a coarse-grained representation of a MOF (comprised of nodes and connecting linkers), and diffusion-based denoising supplies a refined, full-atom representation which can be assessed with various simulation methods.
Similarly, \texttt{SmVAE}~\cite{yao2021inverse} introduces a variational autoencoder that effectively encodes MOF building blocks and stochastically decodes novel MOF structures which are targeted for higher CO\textsubscript{2} capacity. 
Finally, \texttt{GHP-MOFassemble} is a fine-tuned version of the \difflinker{} architecture (a GenAI model originally trained for drug design and discovery) meant for the production of \textit{de novo} linkers for MOFs~\cite{park2024generative}.
For our proposed \mofa{}, we approach GenAI-driven MOF discovery by fine-tuning the \difflinker{} architecture in a high throughput, online learning workflow which periodically re-trains over time.

\subsection{Heterogeneous Computing Workflows}
\label{sec:hetero}

The demands of scientific workflows that couple AI methods with simulation tools~\cite{fox2019learning} has spurred the use of heterogeneous hardware within a single application.
Simulation code may need many CPU and/or GPU cores across many nodes, while AI models are executed most efficiently on specialized accelerators (e.g., GPUs, wafer-scale systems~\cite{cerebras}) with high-bandwidth memory.
Further challenges include flexible routing of results between different components, re-allocation of resources between different task types, and reducing workflow latencies so systems can respond quickly to new information.



Workflow systems enable the expression and execution of applications composed of multiple distinct task types.
The dependencies between tasks are often represented as a directed acyclic graph (DAG) such that the workflow system can optimize placement and execution of tasks in the graph across local or remote resources.
This programming model supports the development of sophisticated computational science applications, and thus, many workflow systems have been developed to meet the needs of the scientific community.
Dask~\cite{rocklin2015dask}, FireWorks~\cite{jain2015fireworks}, Parsl~\cite{babuji19parsl}, Pegasus~\cite{deelman15pegasus}, and Swift~\cite{wilde11swift} all provide mechanisms to express a workflow (tasks and their dependencies) and a runtime for scheduling and dispatching tasks across available resources, such as an HPC cluster.



Solutions for executing an application across heterogeneous resources depend on the physical configuration of those resources.
For example, the challenges faced are different when employing two distinct machines with heterogeneous hardware configurations than when utilizing a single node or a homogeneous cluster.
In the multi-system case, data transfer between remote machines can limit performance and scalability. Additionally, specific tasks may be better suited for a particular hardware configuration, causing bottlenecks in throughput. 
In the single-system case, resource allocation and contention across heterogeneous resources (e.g., CPUs and GPUs) of single node must be considered.
Some workflow systems, such as Parsl, Ray~\cite{moritz2018ray}, and TaskVine~\cite{slydelgado2023taskvine}, support fine-grain allocation of resources such that a subset of CPU cores, memory, or accelerators within a single node can be assigned to a task.
Function-as-a-Service platforms, such as AWS Lambda~\cite{amazonlambda}, Google Cloud Functions~\cite{malawski2020serverless}, and Globus Compute~\cite{chard2020funcx}, support the remote execution of tasks across but lack the fine-grain resource scheduling of workflow systems.


Many science applications have leveraged this increasing hardware heterogeneity and fine-grained workflow systems to improve their computational systems. Workflow management systems have allowed scientists to achieve new computational scales~\cite{al2021exaworks,meyer2023high} across a variety of fields, including virology~\cite{zvyagin2023genslms}, materials science~\cite{guo2022composition}, and astronomy~\cite{villarreal2021extreme}.

\section{\mofa{} Design}
\label{sec:mofa}

Here we provide an abstract formulation for computational MOF design, detail the sequential creation of MOFs from AI generated linkers, and discuss the policies within \mofa{} that enable dynamic MOF generation.

\subsection{Abstract Formulation}

We design \mofa{} as a multi-objective, inverse material design workflow~\cite{zhang2015inverse, dan2020matgan, long2021ccdcgan, kim2020zeogan}.
The goal of multi-objective, inverse material design is to identify materials specified by certain input variables $\Inputs_i$ with properties $P_j$ that satisfy constraints $C_k$.
In \mofa{}, the materials being designed are MOFs, the input variables are linkers and inorganic metals, and the properties and associated constraints are defined as such to produce chemically stable MOFs with high CO\textsubscript{2} adsorption capacity for carbon capture.

MOF properties of MOFs can only be estimated via physical or computational experiments (i.e., there are no formal functional representations for mapping MOFs to desirable properties).
For \mofa{}, we employ computational methods to estimate properties of interest.
Because these properties can only be estimated experimentally, many existing and effective multi-objective optimization methods cannot be directly applied to this problem.
Instead, this problem can be viewed as an \textit{Optimal Experimental Design}~(OED) problem~\cite{huan2024optimal} where we choose actions to perform when aiming to optimize an objective (or set of properties). 
In OED, it is desirable to account for the cost of different actions (e.g., time to perform certain experiments) such as by screening undesirable inputs.
As an example, if given a candidate MOF, it might not be worth performing computationally-intensive estimation of properties if we can more cheaply determine that the candidate MOF is not chemically valid.


We first describe the sequential actions taken---referred to as tasks---to generate a MOF, extending the nomenclature from prior work in steering computational campaigns~\cite{ward2021colmena}.
A \textbf{generator} $\mathcal{G}$ produces a set of linkers $l\in\mathcal{L}$ (e.g., by sampling from a known dataset or generated from an AI model).
Each linker $l$ is \textbf{screened} using an \textbf{assay} $a\in\mathcal{A}$ to estimate a \textbf{property} $P(l)\in\mathcal{P}$ of the linker with the linker being discarded if $P(l)$ does not meet some constraint.
A new MOF $m\in\mathcal{M}$ is \textbf{assembled} from a subset of linkers $\mathcal{L'}\subset\mathcal{L}$.
A series of increasingly expensive but discerning screening steps are applied to each MOF to find a subset of $\mathcal{M}$ with desirable properties.
The specifics of this formal definition are described in \autoref{sec:mofa:sequential}.

While each step in this sequential process could be parallelized via a single program, multiple data~(SPMD) without expression of task dependencies---such an architecture would lead to inefficiencies in staging and sequencing.
Instead, we choose a task parallelism architecture based on the expression of task dependencies. This allows concurrent execution of multiple actions of the same kind, and/or actions of different kinds.
We describe in \autoref{sec:mofa:policies} how \mofa{} defines \textbf{policies} to determine when generation, assembly, and screening should be performed.

\subsection{Sequential MOF Generation}
\label{sec:mofa:sequential}

The following list outlines the discovery pathway of AI-generated MOFs (see \autoref{fig:workflow}), after which we discuss the details of each step (summarized in \autoref{tab:steps}).

\begin{enumerate}
    \item \textbf{Generate linkers}: 
    Use AI model to 
    generate linkers in a form suitable for assembly with pre-selected metal nodes. 
    
    \item \textbf{Process linkers}: 
    Filter linkers without net-zero charge or valid valence number; prepare remainder for assembly.
    
    \item \textbf{Assemble MOFs}: 
    Combine linkers with metal nodes; discard if inter-atomic separations below threshold.

    \item \textbf{Validate structure}:
    Validate MOFs 
    for chemical soundness; compute 
    properties; discard if below thresholds.
    
    \item \textbf{Optimize cells}:
    Further optimize each MOF structure; calculate atomic partial 
    charges of selected MOFs.
    
    \item \textbf{Estimate adsorption}:
    Estimate CO\textsubscript{2} adsorption capacity of 
    successful MOFs and store in database.

    \item \textbf{Retrain}: Retrain \texttt{MOFLinker} using 
    original linker database and linkers of  
    newly screened MOFs.
\end{enumerate}


\begin{table*}
\centering
\caption{
    Details for task types described in \autoref{sec:mofa:sequential}.
    Some tasks employ multiple steps or codes.
    \textit{Remain} is the percent of the original structures (linkers for the first two tasks, MOFs for subsequent tasks) 
    that remain, on average, after the task is performed.
    \textit{Time} in the last column is per structure except for retraining, which is the time to re-train over the entire dataset.
    The resource, remain, and time values are those chosen for or observed during our 450 node run.
    %
}
\label{tab:steps}

\resizebox{\textwidth}{!}{%
\begin{tabular}{
    lllllrr 
}
    \toprule
    
        \textbf{Task} & \textbf{Type} & \textbf{Description} & \textbf{Code} & \textbf{Resource} & \textbf{Remain (\%)} & \textbf{Time (s)} \\ 
    
    \toprule
        
        \multirow{1}{*}{Generate linkers} 
            & Generate & Generate novel linkers & \moflinker{}/\texttt{PyTorch} & 1 GPU & 100.0 & 0.37 \\

    \midrule

        \multirow{1}{*}{Process linkers} 
            & Screen & Screen/optimize linkers & \texttt{RDKit}/\texttt{OpenBabel} & 1 CPU & 22.8 & 0.12 \\

    \midrule\midrule

        \multirow{2}{*}{Assemble MOFs} 
            & Assemble & Connect linkers \& metal clusters& Custom & 1 CPU & 100.00 & 0.46 \\
            & Screen & Check bonds \& atomic distances & \texttt{RDKit} & 1 CPU & 99.90 & 2.56 \\

    \midrule

        \multirow{2}{*}{Validate structure}
            & Screen & Check geometry \& bonds & \texttt{cif2lammps} & 0.5 GPU & 15.20 & 19.98 \\
            & Screen &  Test stability \& porosity & \texttt{LAMMPS} & 0.5 GPU & 8.60 & 204.52 \\

    \midrule

        \multirow{1}{*}{Optimize cells} 
            & Screen & Optimize cell structure & \texttt{CP2K} & 2 nodes & 0.03 & 1517.53 \\

    \midrule

        \multirow{2}{*}{Estimate adsorption} 
            & Screen & Compute partial charges & \texttt{Chargemol} & 1 CPU & 0.03 & 211.78 \\
            & Estimate & Estimate CO\textsubscript{2} adsorption & \texttt{RASPA} & 1 CPU & 0.03 & 1892.89 \\

    \midrule\midrule
    
        \multirow{1}{*}{Retrain}
            & Retrain & Retrain with newly screened MOFs & \moflinker{}/\texttt{PyTorch} & 1 node &  & 96.50 \\
            
    \bottomrule
    
\end{tabular}
}

\end{table*}

 
A key challenge in generating novel molecular structures is ensuring that the model is unaffected by transformations on the molecular structure in the E(3) group (e.g., translation, reflection, rotation, and permutations)~\cite{satorras2021n, batzner2022equivariant}.
\difflinker{}~\cite{igashov2022equivariant} is a state-of-the-art E(3)-equivariant diffusion model originally trained to produce novel molecular structures for drug discovery.
\difflinker{} was trained on the GEOM dataset~\cite{axelrod2022geom} for drug discovery; here, we
fine-tune it with molecular fragments from the hypothetical MOF~(hMOF) dataset~\cite{hmof_2012} to produce a new model, \moflinker{}, that we use to \textbf{generate MOF linkers}.

Since \moflinker{} does not consider hydrogen atoms during generation (it treats them implicitly), to \textbf{process linkers} for assembly we add hydrogen atoms at appropriate locations along the linker and check that their bond lengths and angles are reasonable. \texttt{OpenBabel}~\cite{oboyle_open_2011} is used to determine the bond order and hydrogen atom numbers.
Once hydrogen atoms are added, bond order is determined.
We use the force field \texttt{MMFF}~\cite{1758-2946-3-33} in the \texttt{RDKit}~\cite{landrum2013rdkit} package to reduce stress in 
the linker molecule through energy minimization.
Some linkers may fail these processes and are discarded; remaining linkers are a well-defined molecule with net-zero charge and valid valence number.
Last, the linker anchor parts must be modified before assembly. 
Two types of linker are generated in the workflow: benzenecarboxylic acid (\texttt{BCA}) linker and benzonitrile (\texttt{BZN}) linker.
A \texttt{BCA} linker's carboxylic acid groups are removed, and a dummy atom with element astatine (At) takes the carbon atom's original position; in a \texttt{BZN} 
linker, the nitrogen atoms within its cyano groups are 
identified, and a dummy atom with element francium (Fr) 
is placed 2Å away from each nitrogen atom in the direction 
away from the linker molecule. 
At and Fr are used to label dummy element sites because they are both radioactive and rarely seen in MOFs. 

Thereafter, \textbf{MOF assembly} uses the processed linkers and pre-selected metal nodes to construct new MOFs.
The MOF topology code label is adapted from the Reticular Chemistry Structure Resource (RCSR) database \cite{okeeffe_reticular_2008}, a process that is
automated with custom \texttt{Python} code.
We run several assessments and preparatory tasks using \texttt{RDKit} to ensure that a generated MOF is both reasonable and ready to be simulated with molecular dynamics.
We impute bonds for its given atomic coordinate structure, and determine its SMILES string.
Then, we check that the generated MOF has reasonable bond lengths and angles.
Last, we run a distance-based assessment to ensure that no pair of atoms are overlapping based on a predetermined threshold computed from the experimental database \texttt{OChemDb}~\cite{ochemdb2018}.
If each of these heuristic-based steps pass, the MOF is ready for simulation; if not, it is discarded.

The next step is to \textbf{validate structures} of MOFs with molecular dynamics simulations.
A pre-simulation screen, \texttt{cif2lammps}~\cite{cif2lmp}, is used to ensure that 
atomic structures and chemical bonds are chemically valid within the scope of 
\texttt{UFF4MOF}~\cite{uff4mof,uff4mof2}, a force field used to accelerate the optimization of MOF structures.
Then, a \texttt{LAMMPS}~\cite{lammps} simulation is performed to examine the stability and porous properties of MOFs that have passed prior screens.
For each MOF, 
a 2$\times$2$\times$2 supercell structure is 
equilibrated under a triclinic isothermal-isobaric 
ensemble at $\langle p \rangle = 1\ \textrm{atm}$ 
and $\langle T\rangle = 300\textrm{K}$, such that the cell 
lengths and angles of the MOF structure can be 
equilibrated. These simulations are run for \(10^6\)  
steps with a step size of 0.5~fs. The Linear Lagrangian 
Strain Tensor~(LLST): $S=0.5(e+e^T)$ is calculated for each MOF, 
where $e=R_2R_1^{-1}-I$; $R_1$ and $R_2$ are the unit cell vectors 
for the initial MOF structure and the final MOF structure 
after the \texttt{LAMMPS simulation}; and $I$ is the 3$\times$3 identity 
matrix. 
Eigenvalues of the LLST are calculated, and the 
maximum absolute value of these eigenvalues is chosen as 
the metric to evaluate the lattice distortion before and 
after the simulation.  

\texttt{CP2K} v2024.1 with \texttt{Quickstep}~\cite{Khne2020, qs1996} is then used to \textbf{optimize cells} for each MOF.
Each calculation starts with an initial structure from the prior molecular dynamics simulations,
which is then optimized with a limited number of L-BFGS~\cite{liu1989limited} steps.
Gaussian and Plane Wave~(GPW) method along with 
Perdew–Burke–Ernzerhof~(PBE)~\cite{pbe1996} exchange-correlation 
functional is applied. 
The short range variant of the molecularly optimized basis functions 
with double-zeta valence recommended by Goedecker, Teter, and Hutter 
(DZVP-MOLOPT-SR-GTH)~\cite{gth1996, basis2007} are used. Additionally, 
DFT-D3 of van der Waals correction by Grimme \cite{dftd32010} is added. 

The atomic partial charge is calculated 
using the \texttt{Chargemol} program with the Density Derived 
Electrostatic and Chemical (DDEC6) method~\cite{ddec6p1, ddec6p2}. 
MOFs electronic density in the 3D space is calculated by a single-point energy calculation with \texttt{CP2K}, and the \texttt{Chargemol} program estimates the point charge on each atom that would best fit the calculated electronic density. 
The MOFs failed in atomic partial charge assignment are discarded. 
If the MOFs are successfully assigned with atomic partial charge, their CO$_2$ adsorption value are evaluated using the Grand Canonical
Monte Carlo (GCMC) simulation in \texttt{RASPA}~\cite{raspa2016} (i.e., \textbf{estimate adsorption}).
Specifically, we want to estimate CO\textsubscript{2} capacity at 0.1~bar pressure and 300~K. 
Given the high computational cost and serial execution of GCMC, simulations are conducted under the assumption that MOF structures 
are rigid. The atoms of the MOF structures are assigned 
with Lennard-Jones parameters from the \texttt{UFF4MOF} force field; the default force field model for 
CO\textsubscript{2} within \texttt{RASPA} is used. 
Coulomb forces capture electrostatic interactions in MOFs, 
crucial for gas adsorption. Ewald summation efficiently handles long-range interactions in periodic systems. 
Together, they enhance GCMC, enabling accurate estimates of CO\textsubscript{2} capacity in MOFs. 
Adsorption capacities are stored in the database.  

Periodically, \moflinker{} is \textbf{retrained} on MOFs identified by previous computations.
Retraining starts from the weights learned from pre-training on the hMOF and GEOM datasets~\cite{axelrod2022geom},
and uses a new training set of linkers from as few as 32 and as many as 8192 of the best-performing MOFs yet found during a run.
The training sets are composed of MOFs with high stability ($<$25\% lattice strain) and, at first, only those in the lowest 50\% of lattice strain and then, after 64 gas adsorption calculations have completed, only those with the highest gas adsorption.
Our intent is for the fine-tuned \moflinker{} models to generate linkers similar to those in MOFs with optimal stability and capacity.
Retraining is first performed once 64 stability calculations have completed,
and subsequently after the preceding retraining run has finished and the training set size expands by any amount.
Retraining requires 30--300~seconds, depending on training set size.


\subsection{Workflow Policies}
\label{sec:mofa:policies}

Policies are necessary to dynamically determine what steps to perform at any moment because the sequential screening of entities (linkers and MOFs) means that the number of possible actions and the cost of each action varies throughout the execution of \mofa{}.
\mofa{} utilizes the following policies:
\begin{itemize}
    \item 
    Linkers are continuously generated and processed.
    
    
    \item 
    MOF assembly is performed on the most recently generated linkers
    as soon as enough linkers
    four linkers 
    of each type (\texttt{BCA} and \texttt{BZN}) are available. 
    Assembly runs continuously on one parallel worker for every 256 used for stability calculations. 
    
    \item 
    Computations are performed to assess stability of the most recently assembled MOFs, with a new computation started whenever 
    a stability worker is idle. 
    
    \item 
    Adsorption computations are performed on the most stable MOFs, again with sufficient computations running to maintain full 
    utilization of available workers.
    
    \item 
    \moflinker{} is retrained when at least 64 MOFs with lattice strains below 25\% have been found.
\end{itemize}

The concurrent execution of many steps, with information flowing from one to another and a need to access the most recent (or, in the case of adsorption calculations, the most stable) entity in order to maximize scientific performance, introduces many execution challenges which we discuss in the following section.

\section{Executing \mofa{}}
\label{sec:execution}
\begin{figure*}[t]
    \centering
    \includegraphics[width=\linewidth]{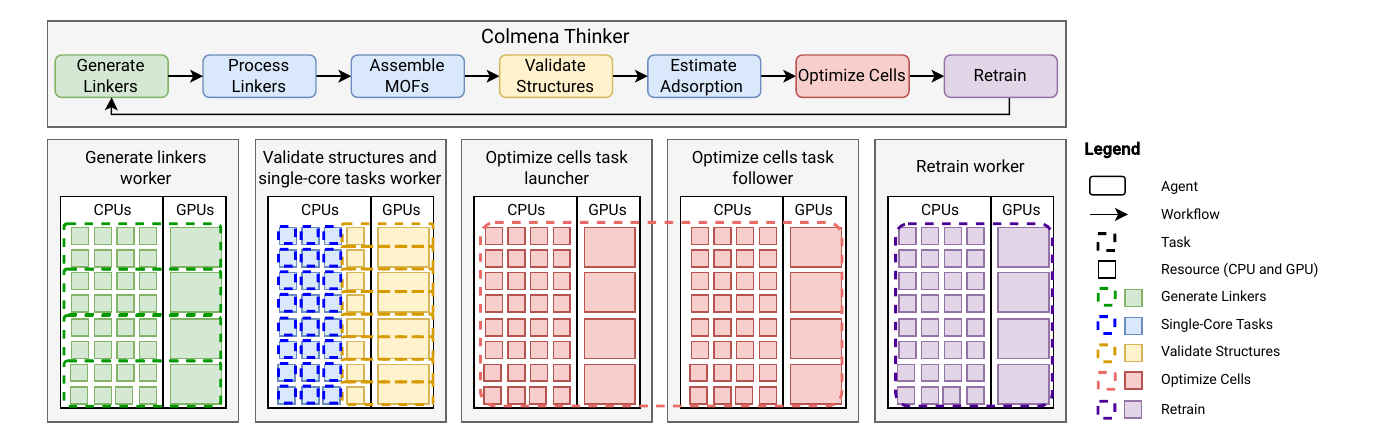}
    \caption{%
       Task and resource allocation in the \mofa{} workflow. The top section shows the \texttt{Colmena} Thinker, containing seven agents (rounded-corner boxes), each corresponding to one of the seven tasks. The bottom section depicts five types of \mofa{} workers, each with a 32-core CPU and four GPUs, with distinct resource allocation schemata for different \mofa{} tasks.
    }
    \label{fig:executor}
\end{figure*}

The dynamic mix of tasks within \mofa{} requires careful attention to policy expression, scheduling, and resource allocation to achieve both high system utilization and efficient and scalable MOF discovery.
We architect \mofa{} to leverage heterogeneous resources and to reduce system latencies (e.g., time to receive a task result) so that new tasks can be determined based on up-to-date information.

\subsection{Policy Expression}

We build \mofa{} on \texttt{Colmena}~\cite{ward2021colmena,ward2023colmena,ward2024colmena}, a Python library for steering simulation ensembles.
In \texttt{Colmena}, a central process, the \emph{Thinker}, executes a set of policies expressed through \emph{agents} that can 
perform actions by submitting \emph{tasks} to a \emph{Task Server}, which manages the remote, asynchronous execution of tasks requested by agents.
Functionally, a task is implemented as a Python function that is executed on a remote process, and agents are threads within the main Thinker process that manage resources, submit tasks, and process task results.

Each of the seven steps described in \autoref{sec:mofa:sequential} are tasks managed by \texttt{Colmena} agents.
A set of agents are implemented to express the \mofa{} policies described in \autoref{sec:mofa:policies}.
For example, one agent is responsible for receiving assembled MOFs, notifying a second agent that resources are available for a new \textbf{assemble MOFs} task, and adding the new MOF to a LIFO queue, to be processed by a third agent that launches a \textbf{validate structures} task when resources are available.
The goal of these policies is to ensure that resources are appropriately allocated between tasks, such as to avoid allocating resources to validate structures when there are insufficient assembled MOFs, and to ensure timely propagation between tasks so that agents make decisions with the most up-to-date information, such as by allocating resources for simulating a more recently created MOF (assuming that MOF quality improves over time).

Prior to this work, \texttt{Colmena} did not have a way to express generator tasks---i.e., tasks that continually yield intermediate data without necessarily returning---which makes it challenging to express \mofa{}'s generative AI tasks.
Thus, we extended \texttt{Colmena} to support Python generator functions that stream intermediate results back to a central process to be consumed and acted upon by an agent.

\subsection{Resource Allocation and Communication}

\mofa{} uses \parsl{} to schedule and execute tasks.
We configure a \parsl{} executor for each resource type and map task types within \texttt{Colmena} to the respective executors.
Rather than submitting a large bag-of-tasks to \parsl{}, \mofa{} only submits tasks when resources allocated to a task type are available.
This choice enables agents to reallocate resources amongst task types depending on the dynamic load across workflow components (e.g., queue lengths).
Notification of a task completion may trigger reallocation of resources and must be done swiftly to maintain high utilization of those resources.
Realizing responsive communication requires reducing costs associated with transmitting results from compute workers to the Thinker and for the Thinker to use them to then decide the next task.

We optimize communication latency by separating workflow control messages (e.g., those used by \texttt{Colmena} and \parsl{}) from result data transfer with ProxyStore~\cite{pauloski2023proxystore, pauloski2024proxystore}.
Sending data through a separate channel speeds the workflow engine's control process, and  decouples actions that involve simply knowing that a task has completed from those that require reading the data.
For example, the Thinker launches the next atomistic simulation as soon as another finishes ($O$(1)~ms latency) and then launches a retraining task once the data from the simulation is processed ($O$(100)~ms latency).

We further accelerate the decision process by distributing the compute-intensive parts of the post-processing (e.g., \textbf{process linkers}) across idle cores on compute nodes.
Distributing tasks to idle cores prevents agents in the \texttt{Thinker} from having to perform post-processing themselves---which would otherwise slow down its ability to respond to new events as quickly.
A final strategy is to process batches of results from inference tasks while others are being run.
We stream results from inference workers to idle cores on other nodes.
ur \texttt{Colmena} agents and Parsl executors, we allocate resources for task type as follows (see \autoref{tab:steps}):
\begin{enumerate}
    \item \textbf{Generate linkers} is performed on a single GPU with a batch size selected to maximize GPU utilization.
    \item \textbf{Process linkers}, \textbf{assemble MOFs}, and \textbf{estimate adsorption} tasks are placed on the idle cores of nodes running \textbf{validate structure} tasks.
    All tasks are isolated by enforcing thread affinity.
    \item \textbf{Validate structure} is configured such that two 
    task invocations share one GPU (via NVIDIA's 
    Multi-Process Service, MPS~\cite{nmps}) but are pinned to different CPUs. 
    \item \textbf{Optimize cells} runs across two dedicated nodes via MPI.
    \item \textbf{Retrain} is performed in a data parallel fashion across all 4 GPUs of a single dedicated node.
\end{enumerate}

The agents cooperate to re-allocate available resources from the pool between tasks types as needed.
A visualization of these and how they are executed across nodes in the HPC cluster can be seen in \autoref{fig:executor}.

\section{Evaluation}

We measure \mofa{} performance along two axes: 
\begin{enumerate*}[label=\textit{(\roman*)}]
    \item computational efficiency on an HPC cluster 
    and 
    \item scientific output.
\end{enumerate*}
We performed experiments on between 32 and 450 nodes of the Argonne Leadership Computing Facility's Polaris Supercomputer, 
an HPE Apollo supercomputer with 560 nodes, each with one AMD EPYC Milan 7543P (32-core, 2.8~GHz) and four 40~GB NVIDIA A100 GPUs.

\subsection{Utilization of Heterogeneous Resources}

\begin{figure}[t]
    \centering
    \includegraphics[width=\linewidth]{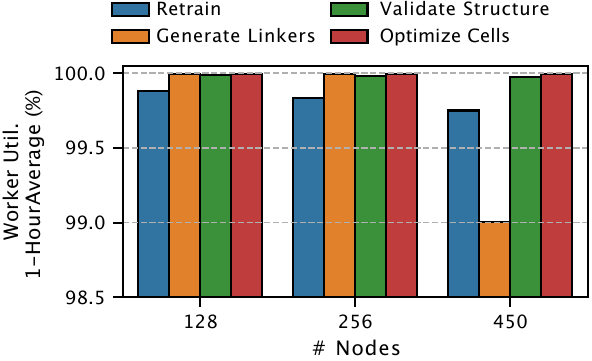}
    \caption{
        Active time of compute nodes on Polaris, 
        as measured by the average time each workflow worker spent processing work over one hour.
        %
        %
    }
    \label{fig:resource_runtime}
\end{figure}
\begin{figure}[t]
    \centering
    \includegraphics[width=\linewidth]{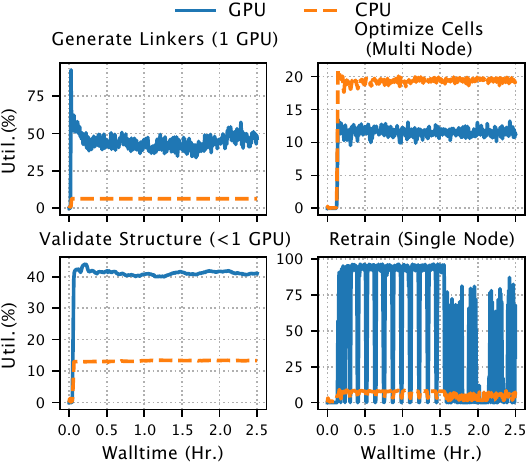}
    \caption{
        \mofa{}'s utilization of Polaris compute nodes as fraction of peak varies with the code running.
        %
        %
    }
    \label{fig:resource_util}
\end{figure}

We first calculate the fraction of time that workers spend doing useful work by analyzing timestamps generated when each worker starts and completes a task.
As shown in \autoref{fig:resource_runtime}, the workers 
for all four task types spend over 99\% of their time executing tasks. 
The workflow makes effective use of every one of the 450 nodes.

Next, we inspect the utilization of hardware within the nodes.
\mofa{} achieves consistent utilization across each type of node during the entire 3-hour run.
We see in \autoref{fig:resource_util} that average GPU and CPU utilization remain constant during a 450-node run for all except the single node workers. 
The single node workers run training tasks, which are large and frequent during the beginning of the run when the application is retraining on any stable MOF and infrequent as the training waits until new gas capacity computations complete.
Only the single node workers maintain near-100\% utilization of the GPU and all have less than full utilization of the CPU. This suggests that we can benefit from further use of NVIDIA's Multi Process Service across all other worker types that under-utilize the GPU (i.e., all but single node workers) in order to execute more tasks per node.

There is room available to offload more post-processing 
from the Thinker to idle CPUs on the compute nodes. 
The validate structure tasks, which are allocated \textless1 GPU, use approximately one quarter of the CPU cores throughout the entire run, 
and generate linkers tasks, which are allocated one GPU, 
use only one eighth.
Thus, there remain idle CPUs and it is possible
to continue our strategy of distributed post-processing across idle 
cores without hindering other tasks.

\subsection{Effect of Scale on Task Throughput}\label{results:throughput}

\begin{figure}[t]
    \centering
    \includegraphics[width=\columnwidth]{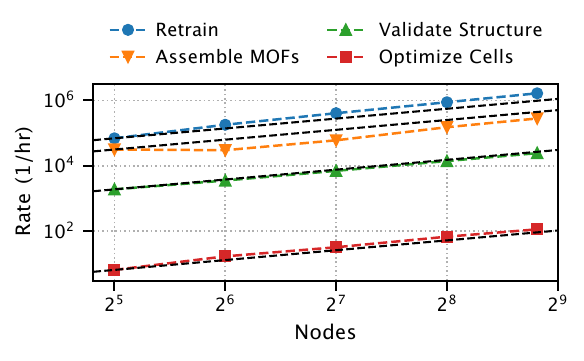}
    \caption{Sustained throughput in tasks per hour for the four main workflow stages as a function of system scale.
    The dashed lines indicate ideal scaling computed from the rates at the smallest node count.}
    \label{fig:throughput-vs-scale-new}
\end{figure}

The rate at which our application evaluates new MOFs increases linearly with scale, as desired.
We measured the throughput by counting the total number of generated linkers, assembled MOFs, structures validated, and cells optimized and estimated then determining a sustained rate using linear regression.
As shown in \autoref{fig:throughput-vs-scale-new}, the throughput for each stage increases linearly from 32 nodes up to a full machine run.

\begin{figure}
    \centering
    \includegraphics[width=\columnwidth]{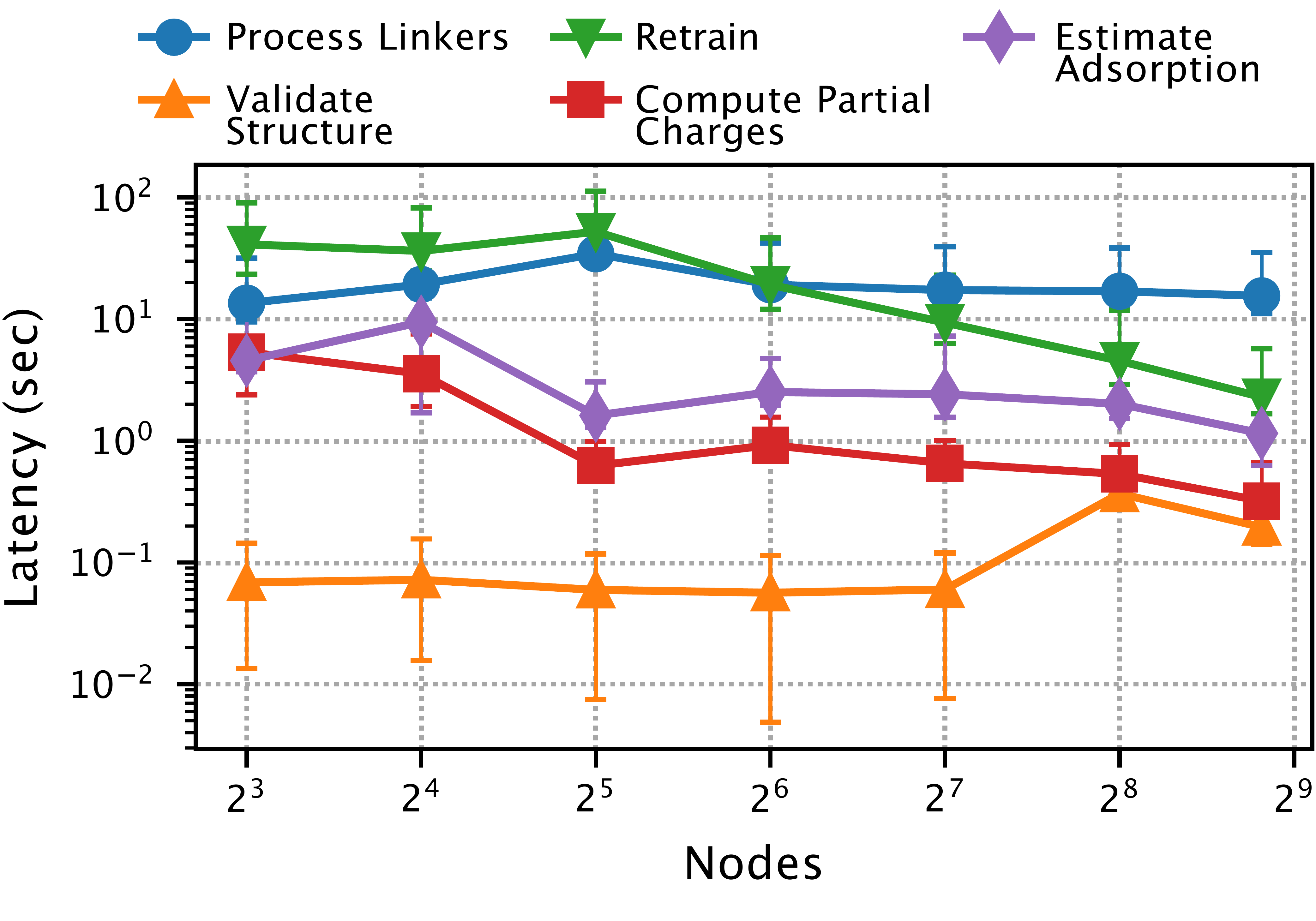}
    \caption{
    Mean and inter-quartile range of key latencies, defined in \autoref{results:throughput}, as a function of node count.
    }
    \label{fig:scaling_latencies}
\end{figure}

The key to scaling in \mofa{} is low inter-stage latencies in the pipeline, because the \mofa{} steering logic only submits enough tasks as available compute resources: results must be processed and new tasks submitted without delay to allow later stages to work on the most up-to-date data.
Consequently, we assess the crucial timings between several stages of the execution plan as a function of scale  to identify potential bottlenecks to further scaling.
As shown in \autoref{fig:scaling_latencies}, we find that the latencies for each of the five critical steps of our application are not degraded by scale.
We assess each below:

\begin{itemize}
    \item Process linkers latency is the time between generating a batch of linkers in a generate linkers task to the Thinker receiving the processed batch from a process linkers task.
    This $O$(10)~s latency is primarily due to the process linkers task runtime; it is constant across node counts, indicating that sufficient CPUs are available for processing. It could be reduced by increasing the parallelism of batch processing.
    
    \item Validate structures latency is the time between a \texttt{LAMMPS} simulation completing and its result being stored in the database.

    \item Retrain latency is the time between finishing retraining a model to that model being used in a generate linkers task.
    Generate linkers tasks complete more frequently at larger scales, leading to a lower latency with scale.
    The latency could be further reduced by adding a mechanism to halt generate linkers tasks when a new model is available.
    \item Compute partial charges latency is the time from an optimize cells task finishing to an estimate adsorption task starting. It remains $\sim$1~s at all scales.
    
    \item Estimate adsorption latency is the time between screening and estimation within estimate adsorption tasks. This also reaches $\sim$1~s at the largest scale.
\end{itemize}


Latencies are kept low by high speed interconnects and messages that for many steps are much less than 1~MB, which do not saturate the network.
The largest tasks (assemble MOFs with 10--40~MB inputs and 1--2~MB outputs, process linkers with 100--500~KB inputs and outputs, and validate structures with 400--600~KB outputs) do require high performance connections to keep communication time in the sub-second range.
We observe $>$1~GB/s transfer rates for many assemble MOF tasks, in particular.
These bandwidth requirements are clearly achievable for Polaris at near full-system scales, and we do not anticipate problems with further scaling.

\subsection{Ability to Find Stable MOFs}

\begin{figure}
    \centering
    \includegraphics[width=\linewidth]{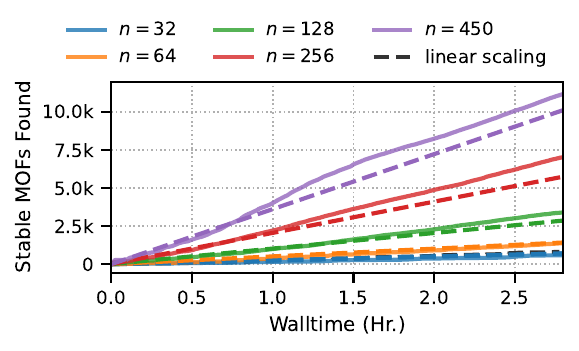}
    \caption{Number of stable MOFs found over time for \mofa{} runs on from 32 to 450 nodes.
    Dashed lines indicated the number of stable MOFs expected by scaling the rate of finding stable MOFs
    over the entire duration of the 32-node run.
    }
    \label{fig:mof-discovery-rate}
\end{figure}

\autoref{fig:mof-discovery-rate} shows the number of stable MOFs (defined as \texttt{LAMMPS} calculation indicating \textless10\% chemical strain) over time.
We attribute the modest increase over time in the rate at which
stable MOFs are generated to repeated retraining of \difflinker{} improving the quality of generated linkers.


We assessed the effect of retraining by repeating the 32-node and 64-node runs with the retraining
portion of the workflow disabled.
The effect of retraining on the stable MOF discovery rate is significant, increasing
the number found at 90 minutes from 133 to 313 on 32 nodes 
and from 393 to 641 on 64 nodes.
The increase in performance is because the fraction of MOFs found to be stable improves.
The fraction of MOFs found to be stable increases from 5 to 11\% when using retraining on 32 nodes
and from 8 to 12\% for 64 nodes.
Learning from intermediate workflow results is clearly beneficial in \mofa{}.

The resources devoted to retraining stay constant at different scales, yet the impact becomes larger.
After 90 minutes, the 450-node run has found 9.7 stable MOFs per node hour expended,
vs.\ 9.5 for the 256-node run and only 6.5 for the 32-node.
We attribute the steady improvement in discovery rate prior to 90 minutes for the 450-node case to more data being gathered, leading to better machine learning models, and---consequently---a more effective \moflinker{} at the same walltime.
(The rate for the 450-node run diminishes after 90 minutes because, unlike the smaller runs, \mofa{} has by then acquired enough
data to switch from retraining based on only stability to a a more stringent combination of stability and gas adsorption capacity.)


\subsection{Novelty and Chemistry Insights of Generated MOFs}

\begin{figure}[t]
    \centering
    \includegraphics[width=\linewidth]{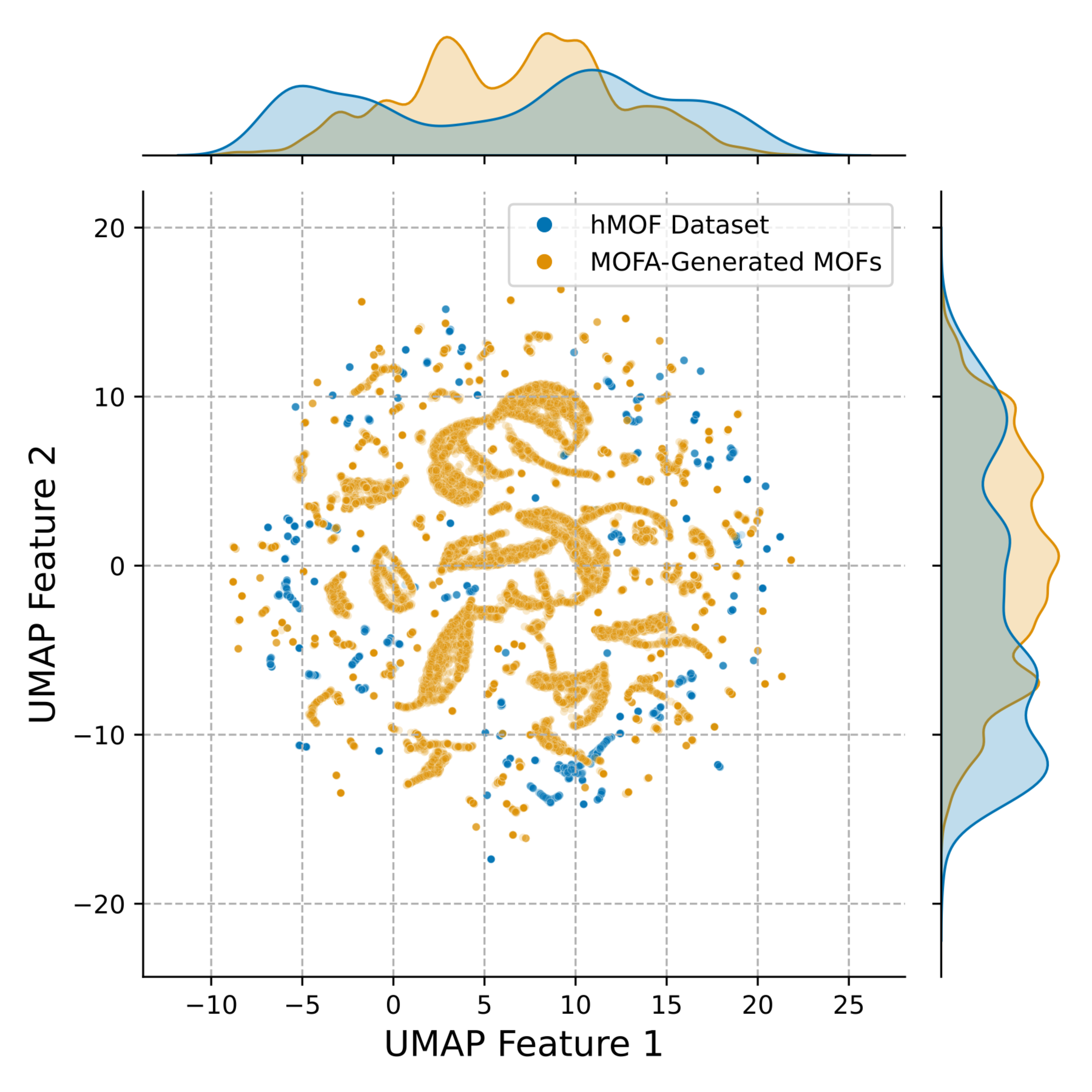}
    \caption{UMAP plot of the diversity of \mofa{}-generated linkers compared to linkers from the hMOF database (represented with an RDKit embedding).
    While some regions of chemical space overlap between hMOF and \mofa{}-generated linkers, the latter explores structures and moieties that differ significantly from those in the original training set---highlighting MOFA's ability to discover new structures within the space of hMOF.
    }
    \label{fig:mofa_umap}
\end{figure}

\begin{figure}[t]
    \centering
    \includegraphics[width=0.9\linewidth]{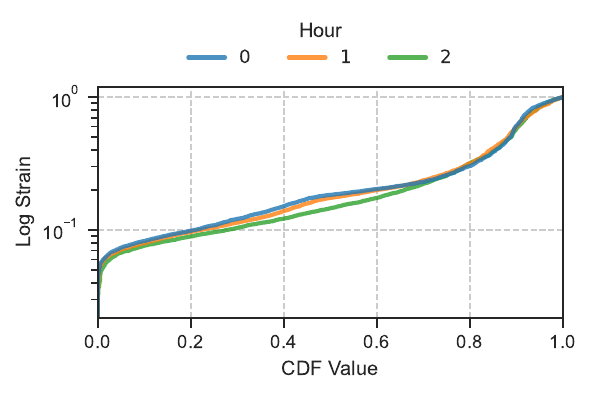}
    \caption{
        The empirical cumulative distribution of the stability of MOFs generated by our 64-node run binned by the hour they were generated. MOF stability (measured by strain) improves over time as the workflow runs. 
    }
    \label{fig:strain_ecdf}
\end{figure}

To evaluate the effectiveness of using \mofa{} to search for MOFs with high stability and CO\textsubscript{2} adsorption, we evaluated the molecules chemical properties over time. In \autoref{fig:strain_ecdf} we compared the cumulative distribution functions~(CDFs) of chemical stability of the generated MOFs for each hour \mofa{} ran. We observe that over time the stability of the MOFs increased, shown by a larger proportion of MOFs having a lower chemical strain. This suggests that our \mofa{} workflow is properly learning to generate MOFs for one of our target objectives.
To understand the chemical novelty of these \mofa{}-generated molecules, in \autoref{fig:mofa_umap} we plot embedding representations of each of the molecules using a UMAP projection based on 38 chemical properties.
While some areas of chemical space were shared between the hMOF database and the \mofa{}-generated linkers, we find that our approach provides candidates that are chemically diverse while sharing important chemical similarities with previously identified successful MOFs.

\begin{figure}
    \centering
    \includegraphics[width=0.75\linewidth]{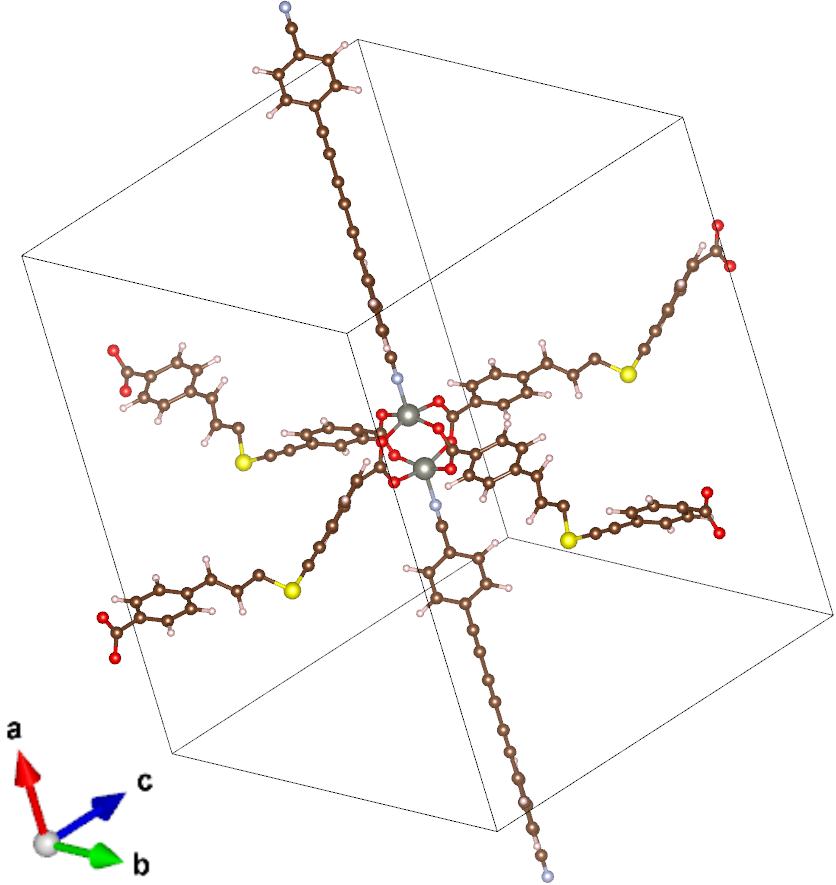}
    \caption{
        The generated MOF with highest CO\textsubscript{2} capacity (4.05 mol/kg at 0.1~bar) produced by a 450-node, 3-hour \mofa{} run on Polaris.  
        Brown: carbon; red: oxygen; white: hydrogen; yellow: sulfur; grey (big): zinc; blue white (small): nitrogen. 
    }
    \label{fig:best_MOF}
\end{figure}

While the chemical stability of the generated MOFs was promising, the goal of \mofa{} is to generate molecules with high CO\textsubscript{2} adsorption. The \texttt{MOFA}-generated MOFs include one with 
CO\textsubscript{2} capacity in the top five of the \texttt{hMOF} dataset, i.e., 
4.05 mol/kg at 0.1~bar: see 
\autoref{fig:best_MOF}. 
Ten other
MOFs produced by the 450-node run also rank in the top 10\% of \texttt{hMOF}, 
with capacities of 1--2 mol/kg at 0.1~bar. 
In brief, with a single 450-node, 3-hour run, \texttt{MOFA} has enabled us to build
a set of novel MOFs with good, and one with very high, CO\textsubscript{2} capacities. 
These results demonstrate \texttt{MOFA}'s capabilities for materials science discovery, and suggest avenues to further improve its performance. 

\section{Impact \& Future Work}
\mofa{}, as a high-throughput generative workflow, presents an opportunity for the discovery of novel MOFs that have a wide array of applications (e.g., reducing greenhouse gas emissions, catalysis);
it is also a compelling application for AI and systems researchers interested in active learning, task scheduling, and data management at large scales.

\subsection{Efficient MOF Discovery}

The promise of MOFs for carbon capture is multifaceted:
\textit{(i)}~their high surface area (from 1000 to \num{10000} m\textsuperscript{2}g\textsuperscript{-1}, exceeding those of traditional 
porous materials such as carbons and zeolites) and 
porosity improve CO\textsubscript{2} adsorption and 
selectivity~\cite{ASEC2023-15374};
\textit{(ii)}~their pore size and shape may be tailored by carefully 
selecting their organic linkers and the connectivity of the 
metal ion clusters;
\textit{(iii)}~they may be designed to maintain their structural integrity 
under harsh environmental conditions;
and
\textit{(iv)}~they can be fabricated at large scale with 
low-cost and simple synthetic methods~\cite{mofs_cost}. 
All these features promote MOFs as a desirable future energy material for carbon 
capture~\cite{LI2019217,SAFAEI2019401}. 
Furthermore, given their unique properties, MOFs have been applied in a number of areas beside carbon capture, including energy storage, catalysis, optoelectronics, and sensing \cite{li2024advances}.
Researchers could therefore adjust the simulation criteria of \mofa{} to search for novel MOFs with applications beyond carbon capture.

However, the many possible clusters and linkers mean that, in principle at least, millions of different MOFs may be created with different properties simply by varying the choice of building blocks. 
Experimental screening of millions of potential MOFs is impractical, and atomistic simulations, like experiments, are too expensive to be used for trial-and-error exploration of the vast MOF chemical design space. 
\mofa{} presents a rigourous approach for efficiently exploring this space by coupling GenAI, high-throughput screening, and atomistic simulations to accelerate the rational discovery of stable, chemically diverse, and high performing MOFs.

We note several other ways in which \mofa{} can be applied for scientific discovery.
We have applied it to create an open source database of high-quality MOFs (URL ommitted for double-blind review).
We aspire also to connect \mofa{} with robotics laboratories that 
synthesize high performing MOFs, and then provide input data regarding experimental synthesizability scores, and cost-effectiveness to manufacture and use such MOFs at scale.

\subsection{Algorithm Research Opportunities}

\mofa{} also presents opportunities for algorithm research.
As described in \autoref{fig:workflow}, \mofa{} employs queue prioritization strategies to determine which structures are used by the next stage in the workflow.
However, many molecule screening procedures (e.g., those in the context of \textit{de novo} drug discovery \cite{wang2024present}) take advantage of adaptive approaches that modify the experiments performed during screening based on new information.
\mofa{} can be readily configured to permit adaptive approaches such as active learning or model-predictive control:
for example, by dynamically re-prioritizing queues with an active learning agent
that optimizes different workflow objectives (e.g., candidate stability, diversity, gas capacity). 
Better algorithms can improve scientific outcomes and/or improve resource allocation, such as by re-prioritizing the DFT simulation queue so 
that computationally expensive experiments are only performed on structures with high predicted gas capacity.

\subsection{Systems Research Opportunities}

\mofa{}'s modular design facilitates the evaluation of different technologies. The configuration of \autoref{sec:execution} works well for our execution environment, but can easily be adapted to support future systems research. The \mofa{} workflow also represents a unique workload due to its complexity and heterogeneity.

The \texttt{Colmena} system that \mofa{} uses for orchestration exposes abstractions including Thinker to Task Server queues for transmitting task requests and streaming results; ProxyStore for intermediate data transfer; and the Task Server for task execution.
Each of these abstractions enables the use of alternate implementations.
For example, researchers can evaluate message broker systems by comparing task latencies in \mofa{}.
ProxyStore enables comparing scalability, latency, and throughput of object stores for intermediate data transfer with its robust plugin system.
Alternate Task Server implementations can be easily created to execute \mofa{} with different execution engines/workflow systems.

\section{Conclusion}

We have presented \mofa{}, an HPC-coupled-generative AI workflow for the accelerated discovery of MOFs for carbon capture.
This workflow leverages heterogeneous computing resources to maximize novel MOF generation through the orchestration of generative-AI, high throughput in-silica screening, and high fidelity atomistic simulations. 
We optimized \mofa{}'s performance by running tasks asynchronously across these workflow modules to maximize resource utilization and throughput of stable MOF generation.
%
%
Once generated, these MOFs were screened for stability and CO\textsubscript{2} capacity. \mofa{} is capable of generating over 100 novel MOFs per hour, and produced 11 promising new candidates for carbon capture in a 450-node three hour run. 
%
We demonstrate the effectiveness of \mofa{} in novel MOF design for carbon capture, while also highlighting the modular nature of our workflow.
It is our hope that \mofa{}'s modular design will
enable future research efforts in distributed systems, as well as in materials science and other science domains
that involve AI and large scale simulations. 




\ifdraft
\else
    \section*{Acknowledgments}
    This work was supported by Laboratory Directed Research and Development (LDRD) funding from Argonne National Laboratory, provided by the Director, Office of Science, of the U.S.\ Department of Energy under Contract No. DE-AC02-06CH11357. 
    This research was partially supported by the Catalyst Design for Decarbonization Center, an Energy Frontier Research Center funded by the U.S. Department of Energy, Office of Science, Basic Energy Sciences under award no.\ DE-SC0023383. 
    This work used resources of the Argonne Leadership Computing Facility, a DOE Office of Science User Facility supported under Contract DE-AC02-06CH11357.  
\fi
   
\bibliographystyle{ieeetr}

\end{document}